# Direct Observation of Reversible Heat Absorption in Li-ion Battery Enabled by Ultra-Sensitive Thermometry

Zhe Cheng[*], Xiaoyang Ji, David G. Cahill

Department of Materials Science and Engineering and Materials Research Laboratory, University of Illinois at Urbana-Champaign, Urbana, IL 61801, USA

[*]Corresponding author: zcheng18@illinois.edu

**Abstract**

The reversible heat in lithium-ion batteries (LIBs) due to entropy change is fundamentally important for understanding the chemical reactions in LIBs and developing proper thermal management strategies. However, the direct measurements of reversible heat are challenging due to the limited temperature resolution of applied thermometry. In this work, by developing an ultra-sensitive thermometry with a differential AC bridge using two thermistors, the noise-equivalent temperature resolution we achieve (± 10 µK) is several orders of magnitude higher than previous thermometry applied on LIBs. We directly observe reversible heat absorption of a LIR2032 coin cell during charging with negligible irreversible heat generation and a linear relation between heat generations and discharging currents. The cell entropy changes determined from the reversible heat agree excellently with those measured from temperature dependent open circuit voltage. Moreover, it is found that the large reversible entropy change can cancel out the irreversible entropy generation at a charging rate as large as C/3.7 and produce a zero-heat-dissipation LIB during charging. Our work significantly contributes to fundamental understanding of the entropy changes and heat generations of the chemical reactions in LIBs, and reveals that reversible heat absorption can be an effective way to cool LIBs during charging.

## Introduction

The control of temperature in lithium-ion batteries (LIBs) is critically important for the performance, reliability, and safety of LIBs.[1,2] Temperature significantly affects the kinetics of charge transport, ion intercalation, and chemical reactions in LIBs. These processes, in turn, determine the heat generation and temperature distributions in LIBs.[3] As the widespread adoption of LIBs in applications such as electric vehicles (EVs), consumer electronics, and grid-scale energy storage, safety concerns such as thermal runaway and poor battery performance at cold temperature hinder the further development of LIBs.[2,4] Additionally, elevated temperature above room temperature, for example 80 C, is beneficial for fast-charging of LIBs.[5] To address these technical challenges, proper thermal management of LIBs is essential to regulate the internal battery temperature.[2,6]

To facilitate advances in designing optimal thermal management strategies for LIBs, greater understanding of heat generation/absorption mechanisms in LIBs is vitally important.[4,7] The thermal effects in LIBs are typically considered to include: reversible heat from cell entropy change, irreversible heat due to ohmic and kinetic losses, heat due to side reactions, and heat of mixing.[4,7,8] The latter two are usually neglected since their contributions are small on normal operating conditions of LIBs.[8] The irreversible heat generation rate can be described as the Joule-heating of the internal resistance.[4]

Compared to irreversible heat effects, the relative importance of reversible heat is controversial in the literature.[7,9-11] The reversible heat of a cell can be expressed as:[11]

$$q_{rev} = T\Delta S I/(nF) , \qquad (1)$$

where $T$ is the temperature, $\Delta S$ is the entropy change of the cell, $I$ is the current, $F$ is the Faraday constant, and $n$ is the number of electrons per reaction ($n = 1$ for LIBs).[11] While charging, the entropy change is negative and therefore the reversible heat is negative. The entropy change of the cell here is the difference of the partial molar entropies for Li of the cathode and anode. By measuring the reversible heat, the entropy change of the cell can be obtained according to Equation (1).

Alternatively, if assuming the chemical Li-ion insertion and extraction which occur at the cathodes and anodes as quasi-static and ignoring the effect of side reactions, the entropy change of the cell is related to the temperature dependence of open circuit voltage (OCV, $U$) according to thermodynamics:[12]

$$\Delta S = F \frac{\partial U}{\partial T}, \qquad (2)$$

The entropy change of a LIB cell is typically derived from Equations (2) in the literature.[4,7,8,11-13] However, the non-linear behavior of the OCV to temperature caused uncertainties in the measurements.[14] The temperature dependent mechanical stresses also possibly have large effects on the entropy changes.[15] Despite of the fundamental importance of entropy change in both batteries and electrochemical heat harvesting,[16-18] the cell entropy changes derived from the OCV measurements have not been experimentally verified previously since the direct measurements of reversible heat without other heat generations are challenging.[4,7]

The relative contributions of different heat generation sources change with charging/discharging rates (C-rates).[4] C-rate is a measure of the rate at which a battery is discharged relative to its capacity. For example, 1C and C/2 mean that the time needed to fully-discharge a battery are 1 hour and 2 hours, respectively. At very low C-rates, the reversible heat dominates the thermal effects in LIBs. In this scenario, the cell entropy change can be measured by directly measuring the reversible heat. The measurement of heat is typically done by a calorimetry which connects the heat source through a thermal resistance to a thermal ground and then measures the temperature rise. However, previous attempts are limited by the temperature resolutions of applied thermometry or energy resolution of applied calorimetry because of the small reversible heat and temperature variation of a cell at very low C-rates.[4,19] The current highest temperature resolutions of thermometry applied on LIBs are usually at the level of tens of mK, which hinders the direct measurements of reversible heat absorption during charging.[4,19] As a result, to our best knowledge, the reversible heat has not been directly measured previously without the effect of irreversible heat generations.[4,7,19] For example, isothermal calorimetry was tried to obtain the entropy change from reversible heat measurements but irreversible heat still dominates the measured heat flux.[20]

In this work, we develop an ultra-sensitive thermometry to measure the temperature variations of a LIR2032 coin cell at very low C-rates. The noise-equivalent temperature resolution is several orders of magnitude higher than those of thermometry applied on LIBs in previous works, which enables us to directly observe the reversible heat absorption in LIB during charging. At C/200, the entropy change can be measured from the reversible heat with negligible irreversible thermal effects (<0.5%). Our work opens the door to a new regime of studying reversible heat effects in LIBs, and contributes to fundamental understanding of the entropy changes and heat generations in LIBs. The C-rate at which the net heat release is zero is also determined, which reveals that reversible heat can be an effective strategy to cool LIBs.

## Results and Discussions

Figure 1 shows the schematic diagram of the thermometry setup and the bridge circuit. As shown in Figures 1(a-b), two thermistors (~10 k$\Omega$) and one resistor heater (~100 $\Omega$) are fixed on two LIR2032 coin cells (40 mAh) seated on a ~3-mm-thick PDMS layer. The PDMS layer is put on an Al plate which is fixed on an optical table. Resistive wires are used to connect these components with the terminal blocks. The Al plates surrounding the thermometry setup are fixed on the optical table by screws and are covered with bubble wrap to suppress heat transfer between the setup and environment from the top and side faces. Figure 1(c) shows the bridge circuit for precise measurements of electrical resistance variations of the thermistor induced by temperature variations of the coin cell. The sine out signal of the lock-in amplifier (SR865) creates a 1-V and 300-Hz AC voltage output as the excitation voltage of the bridge. The thermistor on the left coin cell (marked as $R1_{thermistor}$ in Figure 1(c)) is a matching thermistor, which eliminates the effect of long-time drift of the environment temperature on the measurements. The resistance change of the thermistor on the right coin cell (marked as $R2_{thermistor}$ in Figure 1(c)) induced by the temperature change is obtained by measuring the voltage difference between channels A and B of the lock-in amplifier. The PDMS layer creates a thermal resistance between the coin cell and the Al plate (heat sink) to thermally insulate the coin cell to obtain measurable temperature variations. More details about the experimental setup can be found in the Methods section.

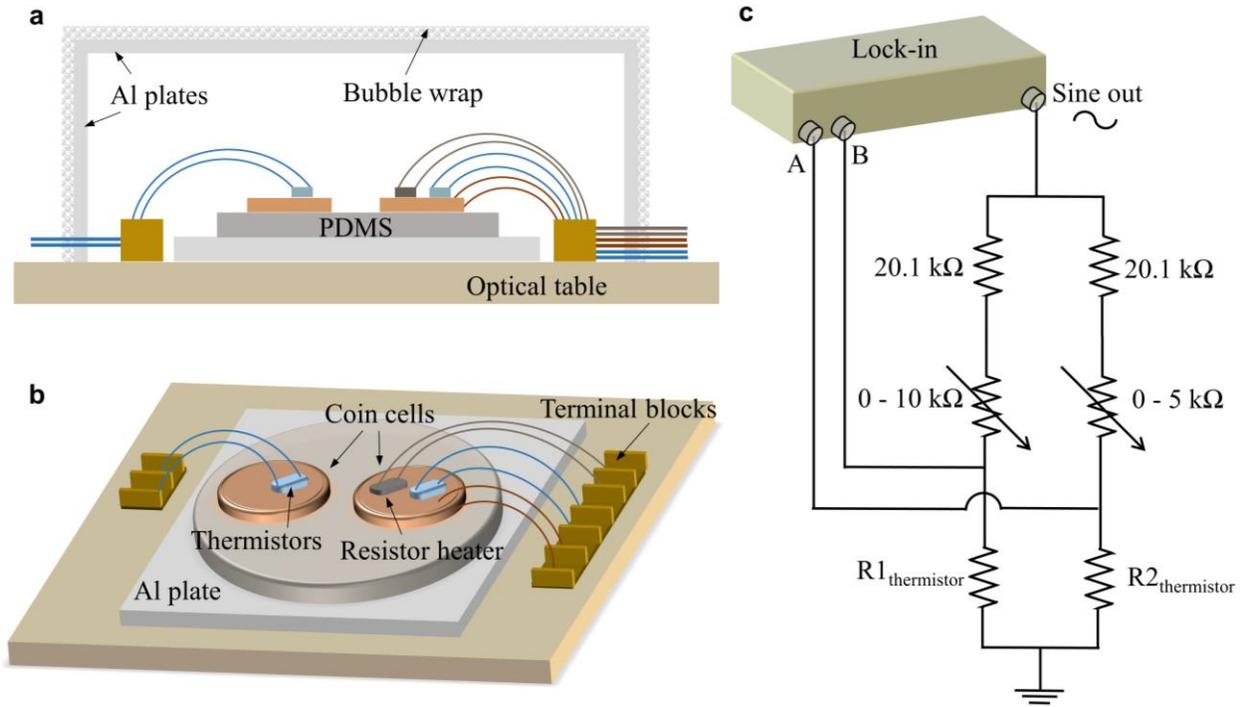

Figure 1. Experimental setup of the ultra-sensitive thermometry. (a) Cross-section image of the experimental setup. Two thermistors and one resistor heater are fixed on the two coin cells seated on a ~3-mm-thick PDMS layer. These components are electrically connected to the fixed terminal blocks. The PDMS layer is put on an Al plate which is seated on an optical table. The top and side Al plates are also fixed on the optical table by screws and covered with bubble wrap to suppress heat transfer between the setup and the environment. (b) Three-dimensional schematic diagram of the thermometry setup. (c) The bridge circuit for precisely measuring the electrical resistance variations of the thermistor due to temperature variations of the coin cell.

The ultra-sensitive thermometry described in Figure 1 has a noise-equivalent temperature resolution of about ±10 μK for ~20 min data collection with a lock-in time constant of 10 s as shown in Figure 2(a). 20 min is the typical time for one reversible heat measurement (10 min charge or discharge and 10 min off). The noise-equivalent temperature resolution is defined as the amplitude of temperature noise variations. Here, the noise is dominated by room temperature fluctuations.

Figure 2(b) shows a comparison of the temperature resolutions and C-rates achieved in this work with the counterparts in the literature. The temperature resolution achieved by the differential AC bridge with two thermistors in this work is several orders of magnitude higher than those of thermometry applied on LIBs in the literature[4,19]: thermocouple,[1,21-24] thermistor,[22] resistance temperature detector (RTD),[6,25-28] fiber Bragg gratings (FBGs),[29-31] infrared image,[22,32-34] liquid-crystal[35]. The C-rates of cells at which the thermal effects in LIBs are measurable are limited by temperature resolutions of applied thermometry and the size of the thermal resistance which also determines the thermal time constant of the system. Therefore, the lower limit of measurable C-rates by our thermometry can be much lower than those in the literature, which we will discuss in details later.

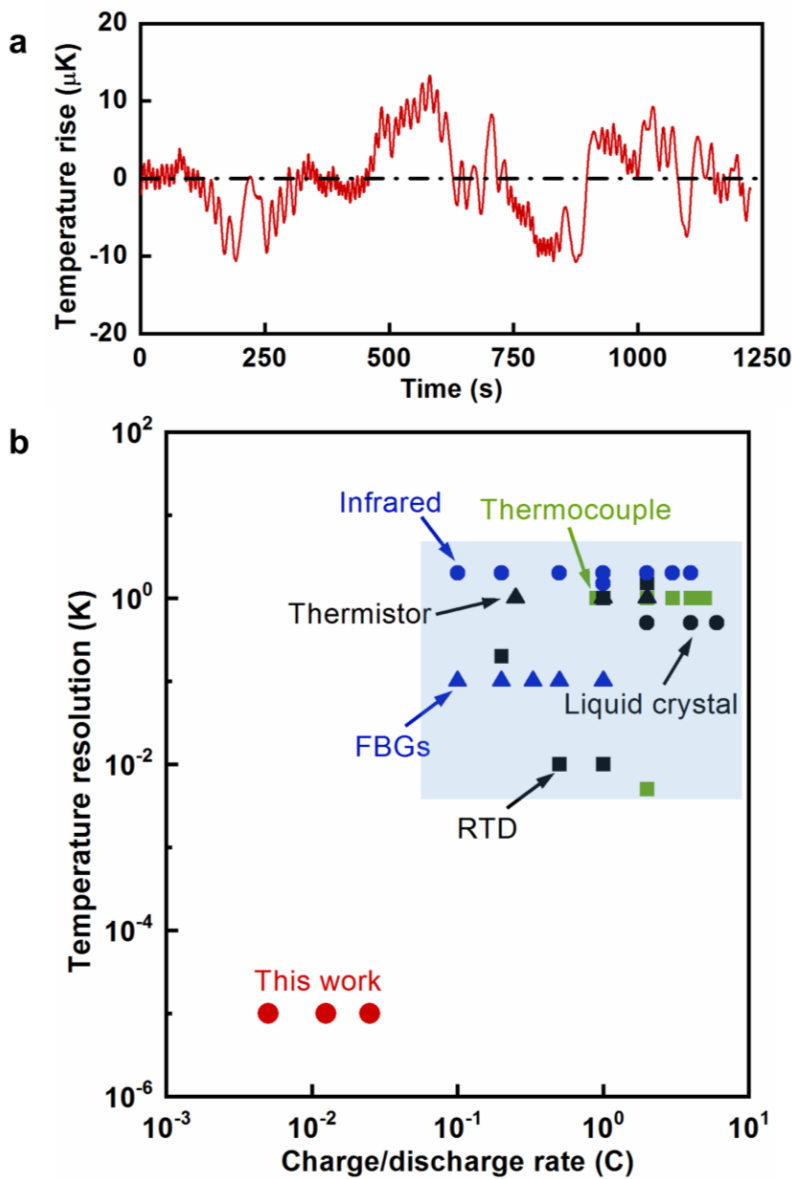

Figure 2. The temperature resolution of thermometry applied on LIBs. (a) Temperature noise of the ultra-sensitive thermometry in this work with ~20 min data collection. The noise-equivalent temperature resolution is 10 μK. (b) A comparison of temperature resolutions and C-rates achieved in this work with the counterparts in the literature[4,19]: thermocouple,[1,21-24] thermistor,[22] resistance temperature detector (RTD),[6,25-28] fiber Bragg gratings (FBGs),[29-31] infrared image,[22,32-34] and liquid-crystal[35].

To charge/discharge the coin cell, a DC current source (Lakeshore 121) is connected to the coin cell. The Aux inputs 1 and 2 of the lock-in amplifier is used to measure the voltage of the coin cell during charging/discharging. A 400-Ω resistor is added into the circuit as a double check of the applied DC current by measuring the voltage across it. The schematic diagram of the circuit can be found in the Supporting Information. The coin cell is discharged for ~10 min, turned off for ~10 min, then charged for ~10 min, and turned off for ~10 min. The applied currents are 1 mA, 0.5 mA, and 0.2 mA, which correspond to C-rates of C/40, C/80, and C/200. It is notable that the ultra-sensitive thermometry in this work enables a new regime of studying thermal effects in LIBs at C-rates orders of magnitude lower than previous works, as shown in Figure 2(b).

The temperature rises of the coin cell are shown in Figure 3. If we define thermal time constant as the time at which the temperature rise reaches 1-1/e of the steady-state temperature rise, the thermal time constant of our setup is about 132 s. Here, e is the Euler's number. If we calculate the time of heat penetrating through the PDMS layer as $d^2/\alpha$, the time is determined as 85 s. Here, $d$ and $\alpha$ are the thickness and thermal diffusivity of the PDMS layer, respectively. The difference in the thermal time constants is attributed to the thermal resistance within the coin cell and the heat capacity of the coin cell. We expect large thermal resistance within the coin cell because the coin cell is composed of a long-strip battery which packs together with polymer separators. The ends of the current collectors are connected with the metal shell of the coin cell. More details about the cell structure can be found in the Supporting Information. The discharging processes are exothermic which result in positive temperature rises while the charging processes are endothermic which result in negative temperature rises (temperature drop). When the C-rate is C/200, the temperature rise and drop are symmetric. In this scenario, the reversible heat dominates the measured heat and all the other thermal effects (irreversible heat generation, heat of mixing, and heat due to side reactions) are negligible (<0.5%). It is notable that this is the first experimental observation of the reversible heat absorption in LIBs without other thermal effects. From this directly measured reversible heat, the entropy change of the coin cell can be obtained, which we will discuss in details later.

As the C-rate increases to C/40, the heating and cooling curves become slightly asymmetric. The temperature rises during discharging are ~11% larger than the temperature drops during charging. The other thermal effects start to take part in the discharging/charging processes, which account for ~5.5% of the total measured heat.

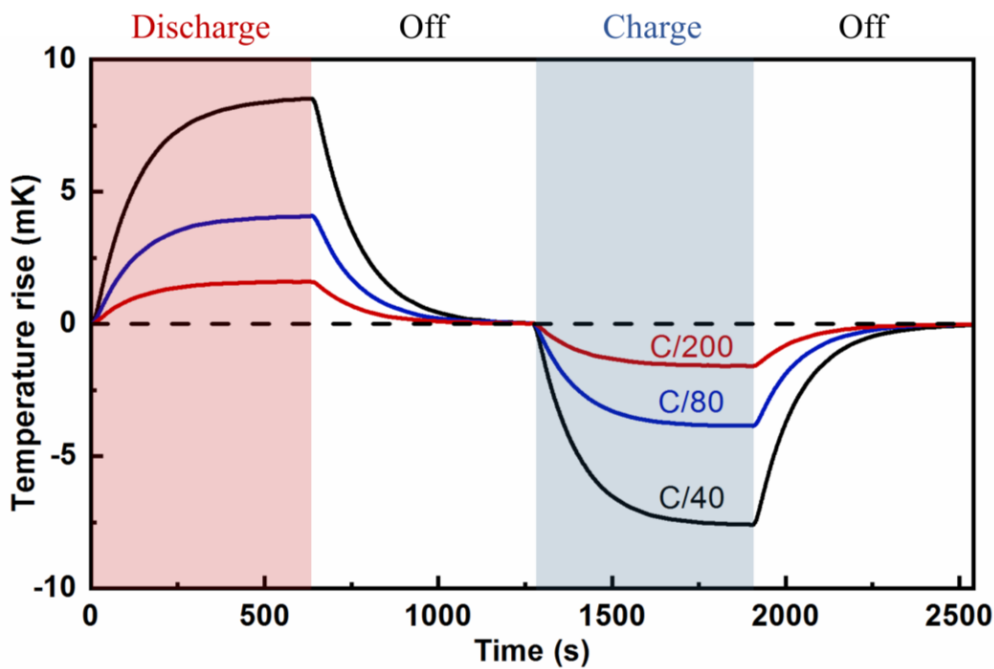

Figure 3. Temperature responses of the coin cell with ~10-min discharging, ~10-min off, ~10-min charging, and ~10-min off for different C-rates (C/40, C/80, and C/200).

As a comparison to the entropy change obtained by the reversible heat, the conventional way to obtain entropy change of LIBs is to measure the temperature dependent OCV.[4,7] To heat up the coin cell, a ~100-Ω resistor heater is fixed on the coin cell and connected with the DC current source. When electrical currents of 25 mA, 35 mA, and 45 mA are applied to heat up the coin cell, the temperature rises and the OCV changes are shown in Figure 4(a). The increase in the temperature of the coin cell results in a decrease in OCV. The relation of the steady-state values of the temperature increases and the OCV decreases are plotted in Figure 4(b). If we do a linear fitting, the slope is -0.51 mV/K, which will be used for the calculations of the entropy change of the coin cell later. Similar measurements are also performed on other state of charge (SOC) of the coin cell. After changing the SOC of the coin cell, at least 4 hours are waited before any measurements in this work.

Similarly, by applying small currents of 0.5 mA and 1 mA through the resistor heater, we can obtain the temperature rises of the coin cell as shown in Figure 4(c). By fitting the applied powers and corresponding temperature rises into a linear relation, the thermal conductance of the coin cell to environment is determined to be 20±1 mW/K. The thermal conductance of the coin cell through the bottom side is the thermal conductance of the 3-mm-thick PDMS layer. The diameter of the coin cell is 20 mm. The thermal conductance of the PDMS layer is estimated to be 15.7 mW/K. The rest of the thermal conductance (4.6 mW/K) should be attributed to the convection and radiation through the top surface of the coin cell. The thermal conductance contributed by the wires connected with the coin cell (two resistive wires for the thermistor, two resistive wires for the resistor heater, and two copper wires for the coin cell) are about 0.25 mW/K if considering them as ultra-long fins on the coin cell surface.

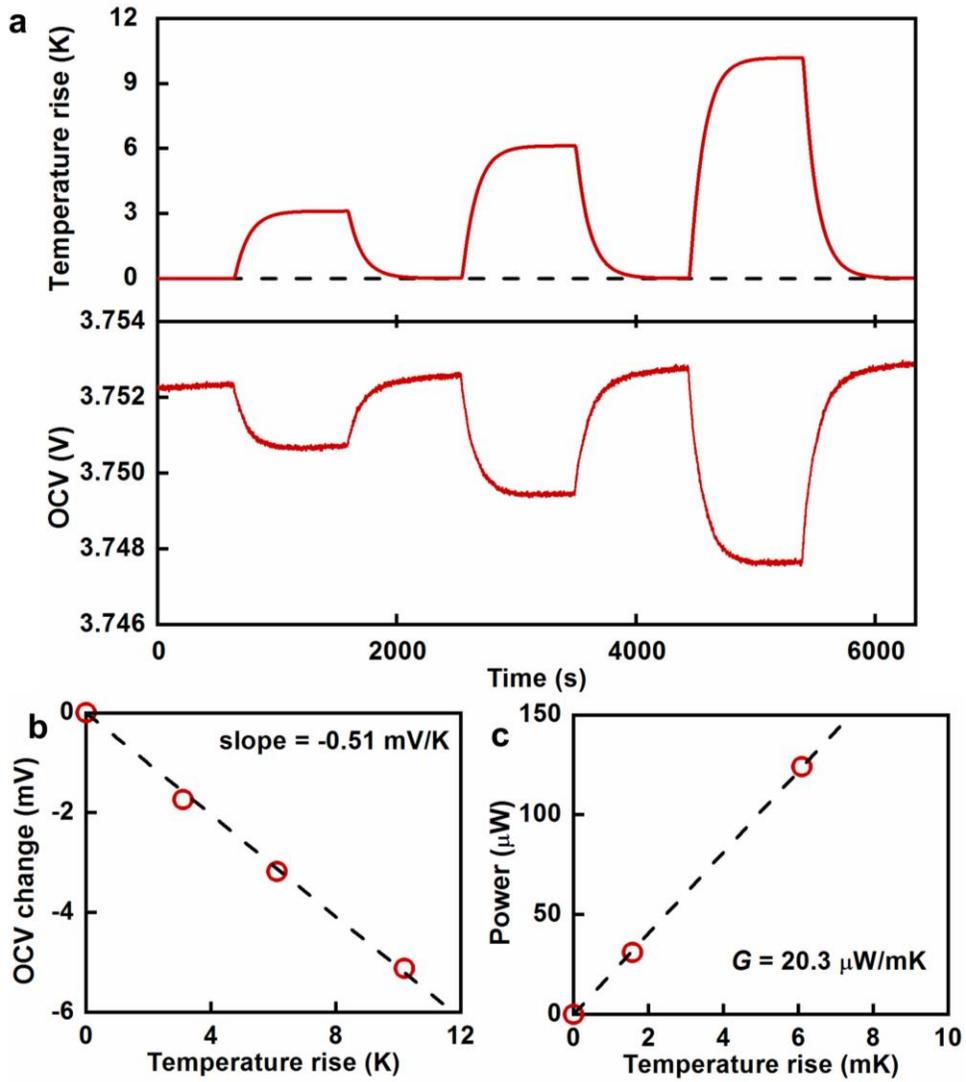

Figure 4. Measurements of entropy change and thermal conductance of the coin cell by heating the coin cell with the attached resistor heater. (a) Temperature rises of the coin cell and temperature dependent OCV. (b) A linear fitting of the temperature rises and the OCV decreases. The slope is determined to be -0.51 mV/K, which will be used to calculate the entropy change later. (c) By measuring the temperature rises as a response of the applied powers, the thermal conductance of the cell to environment is 20 mW/K.

After the thermal conductance of the coin cell is measured, the heat generation of the coin cell can be determined by multiplying the steady-state temperature rises with the measured thermal conductance. The steady-state temperature rises in Figure 3 and the corresponding heat generations are shown in Figure 5(a) as a function of discharge current. The negative values of temperature rises and heat generations correspond to temperature drops and heat absorptions at charge currents. A linear fitting is applied and added into Figure 5(a) as a dashed line. At small discharge currents, the heat generation changes linearly with the currents, which shows that reversible heat dominates the measured heat. When the current increase to 1 mA, the temperature rises start to deviate from the linear relation, showing the existence of the irreversible heat generations.

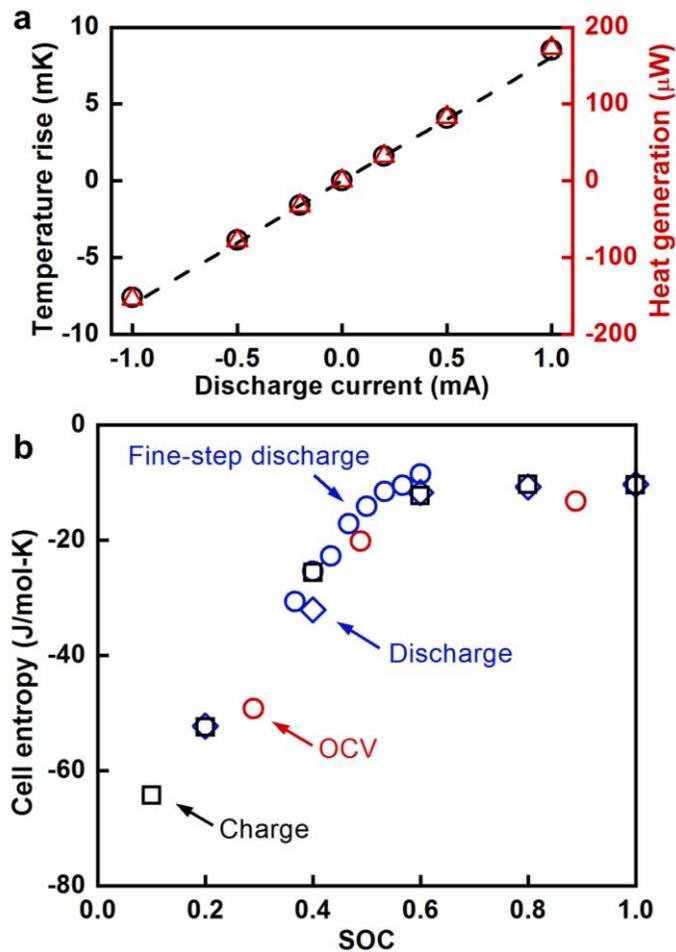

Figure 5. Reversible heat and entropy changes of the coin cell. (a) Temperature rises and heat generations of the coin cell as a function of discharge currents. At small discharge currents, the heat generations change linearly with the currents (the slopes are 8.0 mK/mA and 163 µW/mA), which shows that reversible heat dominates the measured heat. (b) Cell entropy changes of the coin cell measured by reversible heat and temperature dependent OCV as a function of SOC. The "Charge", "Discharge", and "Fine-step discharge" are how we change the SOC to the next SOC after finishing the reversible heat measurements at each SOC. The entropy changes measured by reversible heat and OCV agree excellently.

With the directly measured reversible heat, the entropy changes of the coin cell can be determined according to Equation (1). Additionally, the entropy changes can also be obtained from the OCV data according to Equation (2). Figure 5(b) shows the entropy changes of the coin cell measured by the reversible heat and temperature dependent OCV as a function of SOC. The entropy changes measured by both methods agree excellently at different SOC. Since the changes in mechanical stress with temperature could also possibly change entropy,[15] the approach to measure entropy changes by temperature dependent OCV needs to be verified. Our work is the first verification of this approach and concludes that the possible effect of temperature dependent mechanical stress on the measurements of entropy changes is not important.

To determine the compositions of the cathode and anode materials, energy dispersive X-ray fluorescence (EDXRF) measurements were performed on the cathode and anode materials of the coin cell. More details can be found in the Methods section. The results of EDXRF shows that the cathode materials are $LiCoO_2$ (LCO). The carbon in graphite cannot be detected by EDXRF. Based on the cell voltage and vender information, the anode materials should be graphite. As shown in Figure 5(b), the entropy changes are about -11 J/mol-K and weakly dependent on SOC with SOC from 0.5 to 1 while the absolute values of entropy changes increase with decreasing SOC with SOC from 0.1-0.5. The largest entropy change measured in this work is -64 J/mol-K at SOC of 0.1. The comparison of the OCV-derived entropy changes in this work and those of LCO-graphite cells in the literature can be found in the Supporting Information. Excellent agreement between these entropy change values is achieved, which further confirms that the compositions of our coin cells are LCO-graphite.

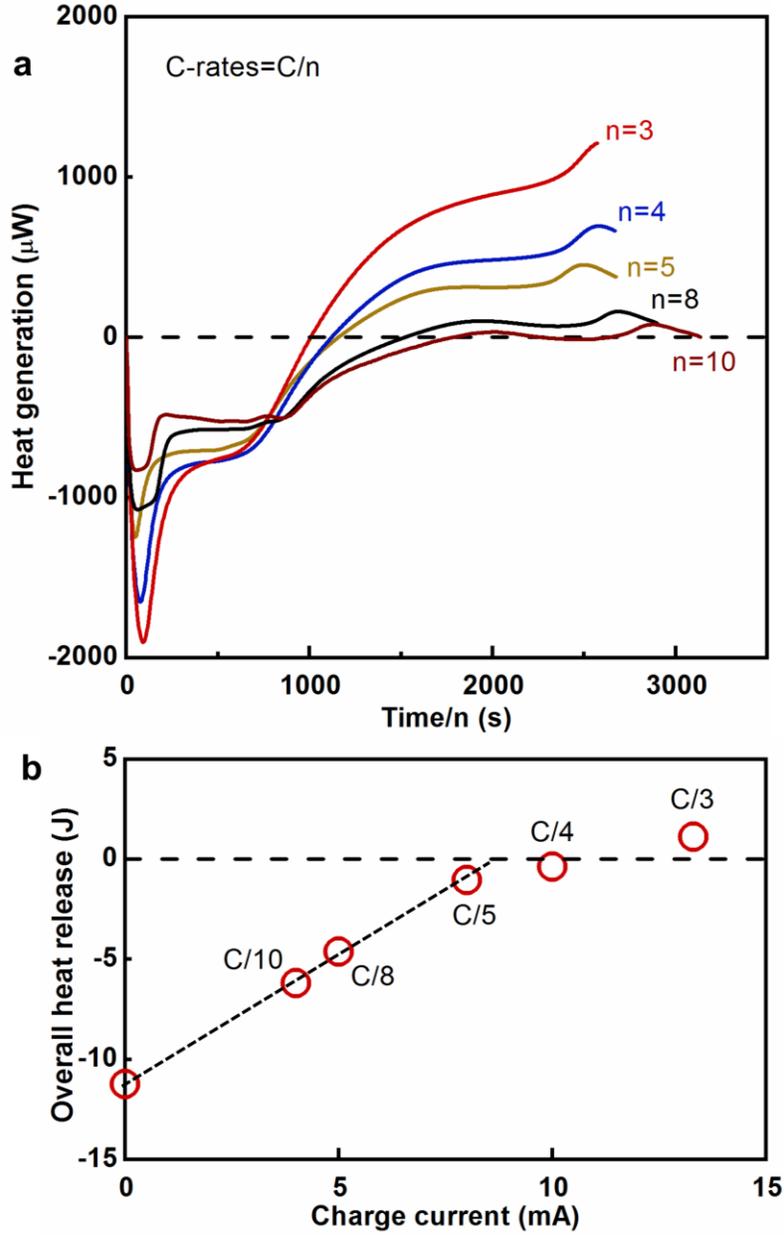

Figure 6. Full-range heat generation of the coin cell. (a) Heat generation of the coin cell during charging with different C-rates (C/3, C/4, C/5, C/8, and C/10). The voltage range for the DC charging processes is 2.5 V-4.1 V. The charging time is divided by n in the figure. (b) The overall generated heat of the coin cell as a function of C-rates. The data point at zero current is the overall reversible heat, which is calculated from entropy changes with SOC from 0.1-0.9.

Figure 6(a) shows the heat generation rates of the coin cell during charging with different C-rates (C/3, C/4, C/5, C/8, and C/10). The voltage range for the DC charging processes is 2.5 V-4.1 V. Significant heat absorptions are observed at the beginning of the charging. The heat absorption rates gradually change to zero before finally turning into heat release. The strong heat absorption is due to the large reversible heat of the coin cell at low SOC. If taking the full-range charging process into consideration, overall heat absorption is observed for the coin cell for C-rates of C/10, C/8, C/5, and C/4. As shown in Figure 6(b), the negative heat release is the net heat absorption of the coin cell. As the increase of the C-rates, the amount of heat absorption decreases. At a C-rate of C/3, the coin cell releases heat for the overall charging process. The C-rate at which the coin cell produce zero net heat to the environment is about C/3.7 according to Figure 6(b). The overall heat release at zero current is total reversible heat, which is calculated based on the entropy changes data with SOC from 0.1-0.9. The dashed line shows an excellent linear relation between the overall heat release and the current because the overall irreversible heat changes with the current linearly. The overall heat release data at C/3 and C/4 is lower than that predicted by the dashed line because the charging times of C/3 and C/4 are shorter than those of C/10 and C/8, as shown in Figure 6(a). The voltage range of all these charging processes is 2.5 V-4.1 V. However, the time it takes to reach 4.1 V is different for each C-rate. The heat generation is high at the time in which the data is not available, leading to the lower overall heat release than that predicted by the dashed line.

## Conclusions

We reported an ultra-sensitive thermometry with a differential AC bridge using two thermistors. The noise-equivalent temperature resolution is ± 10 μK, which is several orders of magnitude higher than those of previous thermometry applied on LIBs. Using this thermometry, reversible heat absorption of a LIR2032 coin cell during charging was directly observed with negligible irreversible heat generations at C/200. The reversible heat was found to be a linear function of the charge/discharge currents. Furthermore, the entropy changes of the cell were directly obtained from the measured reversible heat. For comparison, we also measured the temperature dependent OCV to obtain entropy changes and found excellent agreement. It is notable that this is the first verification of the approach of obtaining entropy changes by measuring temperature dependent OCV since the temperature dependent mechanical stress could also affect entropy changes. We found that the large reversible entropy change can cancel out the irreversible entropy generation at a charging rate as large as C/3.7 and produce zero net heat during charging the coin cell. Our work not only significantly contributes to fundamental understanding of the entropy changes and heat generations of the chemical reactions in LIBs, but also reveals that reversible heat absorption can be an effective way to cool LIBs during charging.

## Methods

**Experimental components:** The coin cells are purchased from Amazon (CT-Energy, LIR2032, 40 mAh). The thermistors are purchased from Digi-Key (Murata, NCP18XH103D03RB, ~ 10 k$\Omega$) and the temperature coefficient of resistance is -429.4 $\Omega$/K at room temperature (~22 C). The thermistors are 1.60 mm in length and 0.80 mm in width. The resistor heater is purchased from Digi-Key (Susumu, RG20N100WCT, ~100 $\Omega$). The resistive wire is nichrome 80-36 AWG Gauge Spools 300 feet with a resistance of 35.6 $\Omega$/feet (30A-T197 from FogsLord). The PDMS is purchased from Amazon (Sylgard 184, 24236-10). 10g silicone elastomer and 1g curing agent were mixed together to form a 3-mm-thick disk. The curing process was finished at room temperature overnight before the PDMS layer was transferred to the Al plate.

**Wire connections**: To suppress the heat conduction via the wires from the thermistors, resistive wires are used to connect the two thermistors. The thermal conductance contributed by a resistive wire connected to the coin cell is about 11 μW/K. To keep good thermal contact between the resistor heater and thermistors with the coin cell, the soldered ends of these components need to be flat and at the same surface with the backside surface of these components. First, the resistor heater and thermistors are fixed on a glass slide by superglue, which makes sure that the backside surfaces of these components have good contact with the flat surface of the glass slide. Then the ends of these components are soldered with resistive wires. After that, the resistor heater and thermistors are removed from the glass slide and adhered to the top surface of the coin cell by superglue. The superglue also acts as an electrically insulating layer between the resistor heater and thermistors and the top surface of the coin cell. Epoxy glue was used to further fix the resistor heater and thermistors on the top surface of the coin cell. Copper wires were used to connect the two electrodes to the terminal blocks. The ends of the copper wires were soldered with tin to obtain a large surface to contact with the coin cell. Tweezers were used to press the end of the copper wire to obtain good electrical contact. Then epoxy was used to fix the tin-extended end of the copper wire and silver paste was used to further enhance the good electrical contacts between the copper wires and the coin cell. Two coin cells were soldered with copper wires to keep the thermal conductance of the coin cells the same even though we only charge/discharge one coin cell. More details about the wire connections can be found in the Supporting Information.

**Temperature measurements**: The room-temperature electrical resistance of the two thermistors is about 11.3 kΩ. The two coin cells are thermally connected with the heat sink (Al plate) via a thermal resistance (a uniform PDMS layer). When the environment temperature fluctuates, the temperatures of the two coin cells change similarly. The corresponding electrical resistances of the two thermistors change similarly and the bridge voltage is still balanced. The measured voltage by the lock-in does not change, which attenuates the effects of environment temperature drift. The electrical resistances of the two thermistors are not exactly the same so two tunable resistors were added on the other two arms of the bridge to match the resistance ratio of the two thermistors. The two tunable resistors are 0-10 kΩ and 0-5 kΩ but only tens of Ω is typically used on either arm since the difference of the resistances of the two thermistors is small (tens of Ω). The lock-in amplifier was used to measure the voltage difference of the two thermistors at the same frequency as the excitation source. The in-phase voltage signal was used to calculate the temperature variations of the coin cell. For the choosing of the excitation voltage of the bridge, we need to obtain high temperature resolution for our measurements and also avoid generating too much Joule-heat. In this work, we used 1 V as our excitation voltage. Further increase in excitation voltage does not increase the temperature resolution much but induces additional Joule-heating. Here, 1 μV change in the measured voltage corresponds to 0.114 mK change in the temperature of the coin cell. The derivations can be found in the Supporting Information. The heat generation of the thermistors under 1-V excitation voltage is 5.5 μW, leading to a temperature rise of 0.27 mK for the coin cell. Both thermistors have similar heating and steady-state temperature rises. As a result, the bridge is still balanced and the bridge voltage is not affected. Additionally, when measuring the temperature dependent OCV, the temperature rises are several Kelvins. The non-linear resistance-temperature relation of the thermistor is used to calculate the temperature, unlike the measurements of small temperature rises in which a linear approximation of the resistance-temperature relation is used. More details about data processing can be found in the Supporting Information.

**EDXRF measurements**: EDXRF measurements were carried out to determine the compositions of the cathode and anode materials by Shimadzu EDX-7000 spectrometer with the X-ray beam from an Rh anode. After the cathode and anode materials were scratched from the Al and copper foil current collectors, they were collected to the sample tubes. The bottom of sample tube was sealed by a 4-μm-thick polypropylene film. The X-ray beam collimated into a diameter of 3 mm was incident onto the sample through the polypropylene window. The fluorescence X-rays were collected by a Silicon drift detector. PCEDX Navi software was used to control the scan running and analyze the spectra. The quantitative scan shows that Co dominates in the cathode materials. The carbon in graphite cannot be detected by EDXRF because EDXRF is not sensitive to light atoms such as Li, carbon, and oxygen.

## Acknowledgements

We would like to acknowledge the financial support from US Army CERL W9132T-19-2-0008.

## Competing interests

The authors declare no competing interest.

## Data availability

The data that support the findings of this study are available from the corresponding authors upon reasonable request.